# Research on the Influence of Underwater Environment on the Dynamic Performance of the Mechanical Leg of a Deep-sea Crawling and Swimming Robot

Lihui Liao
School of Mechanical Science & Engineering
Huazhong University of Science and Technology
Wuhan, China
liaolihuiysu@163.com

Baoren Li
School of Mechanical Science & Engineering
Huazhong University of Science and Technology
Wuhan, China
lbr@hust.edu.cn

Dijia Zhang
School of Mechanical Science & Engineering
Huazhong University of Science and Technology
Wuhan, China
zhangdijia323@163.com

Luping Gao
School of Mechanical Science & Engineering
Huazhong University of Science and Technology
Wuhan, China
15346787206@163.com

Mboul éNgwa
School of Mechanical Science & Engineering
Huazhong University of Science and Technology
Wuhan, China
ngwa_m@yahoo.fr

Jingmin Du
School of Mechanical Science & Engineering
Huazhong University of Science and Technology
Wuhan, China
hustdjm@hust.edu.cn

*Abstract*—The performance of underwater crawling and adjustment of the body posture for underwater manipulating of the deep-sea crawling and swimming robot (DCSR) is directly influenced by the dynamic performance of the underwater mechanical legs (UWML), as it serves as the executive mechanism of the DCSR. Compared with the mechanical legs of legged robots working on land, the UWML of the DCSR not only possesses the characteristics of the land used mechanical legs, but is also affected by the influence of the deep-sea underwater working environment (i.e., the hydrodynamic force, viscous resistance and dynamic seal resistance). To reduce these influence, firstly, the hydrodynamic force of the UWML were researched based on theory and experiment, and the hydrodynamic model was established with the fitted hydrodynamic parameters. Secondly, the oil viscous resistance and the dynamic seal resistance were studied experimentally, and the change laws of both with respect to the joint speed and the ambient pressure (depth of operation) were obtained. The results provide a basis for the subsequent research on the structure optimization and high performance control of the UWML and DCSR.

## I. Introduction

With the increasing scarcity of land resources, the pace of human developing and utilization of marine resources is accelerating, and gradually moving from shallow waters to deep sea. In recent years, with the continuous development of robot technology, robots have been increasingly applied in the field of ocean engineering, and deep-sea robots are therefore attracting more and more attention and research. Compared with wheeled and tracked robots, legged robots have flexible mobility, powerful obstacle avoidance ability, and excellent terrain adaptability [1], thus becoming a hot research topic in recent years [2-5]. As one type of legged robots, deep-sea crawling and swimming robot integrate the swimming function of underwater robots and the walking function of legged robots. Not only they can swim and crawl on the seabed, but also use their manipulators for underwater operations, which make them becoming an important research equipment for ocean engineering.

Seabed crawling, underwater patrolling and deep-sea manipulating are the three basic functions of the DCSR when performing underwater tasks. As the executive mechanism for the crawling and body posture adjustment of the robot during underwater operations, the dynamic performance of the UWML is directly related to the performance of the crawling movement and deep-sea manipulating. As a multi joint serial mechanism, the UWML is a complex dynamic system with multiple inputs and outputs, strong coupling, and nonlinearity [6]. Due to the influence of the special underwater working environment, the UWML exhibits three particular characteristics compared with the mechanical legs used on land: (1) The motion of the UWML will be disturbed by the hydrodynamic force [7]; (2) The pressured compensation oil will make the high speed rotating parts inside the UWML (such as motors, etc.) to be affected by the oil stirring viscous resistance [8-10]; (3) The high-pressure of the deep-sea environment will also lead to the sharply increasing of the dynamic seal resistance on the joint output shaft [11]. Therefore, the UWML will be subjected to the combined effects of the hydrodynamic force, viscosity resistance, and dynamic seal resistance during underwater working, and the dynamic model of the UWML tends to be more complex, not only exhibiting strong nonlinear characteristics (i.e., structure nonlinearity, friction resistance and oil viscous resistance nonlinearity, etc.), but also adding lots of uncertainties (such as parameter uncertainty, external hydrodynamic disturbance uncertainty, etc.). As a result, the motion performance of the UWML is deteriorated and the control difficulty is increased. To improve the control performance, it is necessary to found the change laws of the hydrodynamic force, viscous resistance, and dynamic seal resistance of the UWML.

Researchers have studied the change laws of the hydrodynamic force of several types of robotic arms. References [12,13] established the hydrodynamic models of underwater manipulators based on the Morison's formula with empirical values of the hydrodynamic coefficients. Reference [14] obtained the added mass force and drag force coefficients of a double link arm by slice theory and experimental testing. Reference [15] combined flow visualization, theoretical analysis, and torque sensor measurements to investigate the hydrodynamic force of a single-joint underwater manipulator, where the added mass force and drag force coefficients were associated with the system's motion states. Reference [16] studied the transient hydrodynamic coefficients of a single DOF underwater manipulator with circular cross-section by numerical simulation methods and extended it to the transient hydrodynamic coefficients calculation of a square cross-section single DOF underwater manipulator [17], from which the drag coefficient and added mass force coefficient were related to the COM motion distance.

For the change laws of the viscous resistance caused by stirring oil when joint motor rotating, some researches has been conducted. Reference [18] studied the viscous resistance of fluid in the annular gap between coaxial cylinder using laminar flow theory, and summarized the relationship between viscous resistance and Taylor coefficient. Reference [19] simulated the viscous resistance of fluid in the air gap of an underwater oil-filled brushless DC motor with a stationary rotor through the computational CFD method, and analyzed its variation laws with respect to radial clearance size, while reference [20] experimentally investigated the change laws of the viscous resistance of fluid in the air gap of an underwater oil-filled brushless DC motor with respect to pressure and temperature, and proposed a practical method for estimating viscous resistance based on the selected viscous resistance model and the oil's characteristics. Reference [9] conducted experimental tests to analysis the effects of viscous resistance of compensating oil on the dynamic performance and load capacity of a deep-sea electric manipulator. Based on the existing theoretical models, a new viscous resistance model was proposed that considers the effects of surface grooves of the rotor and stator, and a method was proposed to optimize the clearance geometry by filling the surface grooves of the rotor and the stator with epoxy resin in order to reduce the viscosity losses.

For the change laws of the dynamic seal resistance of the joints in underwater robotic arm, there have been few studies. Reference [21] compensated the dynamic seal resistance of joints for compliance control of a hexapod robot leg with testing data. However, the change laws of the dynamic seal resistance under different depth had not been studied. Reference [22] experimentally studied the dynamic seal and pressure compensation performance of a underwater motor, but focused on its reliability and overall operational efficiency, without studying the variation laws of the dynamic seal within motor.

Summarizing the above analysis, it can be seen that the current research on the influence of underwater environment on the dynamic performance of robotic arms like the UWML has not concluded general rules, and the studies are generally conducted on the one or two aspects by experiments or semi-theoretical and semi-experimental methods for specific applications. To reduce the disturbance of the underwater environment on the dynamic performance of the UWML, this paper firstly established the hydrodynamic model with fitted hydrodynamic parameters through a semi-theoretical and semi-experimental method, and then analyzed the proportion of each hydrodynamic component. Secondly, the viscous resistance and dynamic seal resistance of the joint of the UWML were studied experimentally, and the change laws of both with respect to the joint speed and working pressure were tested, as well as the effect of both on joint efficiency.

This paper is organized as follows. Section 2 introduces the structure of the UWML. Section 3 presents hydrodynamic modeling, hydrodynamic force testing and hydrodynamic parameters fitted procedure. Section 4 carries out experimental studies of the viscous resistance and the dynamic seal resistance of watertight joint of the UWML, and some conclusions are summarized in Section 5.

## II. STRUCTURE OF THE UWML

The UWML of the DSCR is a 3-DOF serial mechanism designed based on the bionic prototype of the crab leg, as shown in Fig. 1. The UWML mainly consists of six parts: a base (including electronic cabin), a thigh link (link 2), a calf link (link 3), a hip yaw joint (joint 1), a hip roll joint (joint 2), and a knee roll joint (joint 3). The unique working environment of the deep sea brings a series of new requirements for the design of the UWML, such as water tightness, pressure resistance, corrosion resistance, etc. Based on these special design requirements, the UWML studied in this paper possesses five structural features, namely, water tightness, modularity, oil pressure compensation, hollow cable layout, and dry-wet separation. The structural characteristics and the special working environment create the differences of the UWML from land-used mechanical legs, and both are directly related to the hydrodynamic, viscous resistance, and dynamic seal resistance of the UWML. The structure of links and joints affects the hydrodynamic force, while the compensation oil pressure and the structure of the high-speed rotating parts inside the joint affect the viscous resistance, and the working environment pressure and the structure of the joint axis affect the dynamic seal resistance.

## III. HYDRODNAMIC FORCE OF THE UWML

Due to the viscosity and buoyancy of water, the UWML will be subject to hydrodynamic force and buoyancy when moving in water, and exhibit different dynamic characteristics compared with those used on land. The magnitude of the hydrodynamic force and buoyancy is not only related to the properties of water, but also to the shape, structure, and motion states of the UWML. According to the theory of fluid mechanics, when a single object moves in water, the force from water is:

$$\frac{dF}{dl} = \frac{dF_d}{dl} + \frac{dF_m}{dl} + \frac{dF_l}{dl} + \frac{dF_f}{dl} \quad (1)$$

where $F_d$ is the drag force, $F_m$ the added mass force, $F_l$ the lift force, and $F_f$ the buoyancy. $F_d$ and $F_l$ are generated from the pressure and friction of water acting on the object during relative motion, respectively, which act on the direction of the inflow velocity and perpendicular to the inflow velocity. When the object has no wing-like structure, $F_l$ can be neglected. In addition, when object is accelerated in water, there is also an active force that makes the water around the object to accelerate. As a result, the object is subjected to the reaction force of water, which is the additional mass force $F_m$.

In 1950, Morison conducted a study on the hydrodynamic force of an object moving in water, which primarily considered the effects of the water drag force $F_d$ and added mass force $F_m$. Based on the research results, He proposed the Morison formula for calculating the hydrodynamic force [23]:

$$d\boldsymbol{F} = d\boldsymbol{F}_d + d\boldsymbol{F}_m = \frac{1}{2}\rho C_{\mathrm{d}} D \parallel \boldsymbol{v} \parallel \boldsymbol{v} dl \\ + \rho C_{\mathrm{m}} A \dot{\boldsymbol{v}} dl \quad (2)$$

where $\rho$ is the water density, $C_d$ the drag force coefficient, $D$ the equivalent diameter of the object, $\boldsymbol{v}$ the velocity of the inflow, $C_m$ the added mass coefficient, $A$ the projected area of the object in the direction perpendicular to the flow velocity $\boldsymbol{v}$, and $\dot{\boldsymbol{v}}$ the acceleration of the inflow. Research shows that both $C_d$ and $C_m$ are related to the Keulegan-Carpenter (KC) number, Reynolds number $Re$, object shape and structure, as well as the orientation of the flow field.

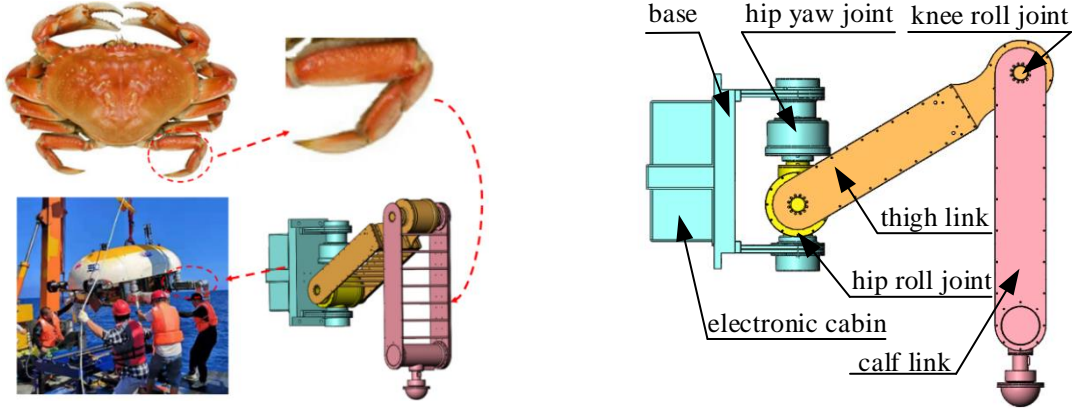

Figure 1 The structure of the UWML

Currently, due to the immaturity of relevant theories, it is impossible to acquire their accurate values based on theory. Therefore, their values are generally obtained through experimental testing for engineering application [7].

As shown in Fig. 2, a square link (with a cross section of height $D_1$ and width $D_2$) with a length of $l$ is submerged in water and rotates around the $Z$ axis of the coordinate system $\{\,O\,\}$. The water resistance acting on the link can be decomposed into a normal resistance force $dF_d$ and a tangential resistance force $dF_t$. The direction of $dF_t$ is tangent to the surface of the link and its value is generally small, so it can be ignored in the analysis. The direction of $dF_d$ is perpendicular to the surface of the link, which has a greater effect on the motion of the link, so it cannot be ignored. According to the slice theory, a slice $dx$ can be taken at a position $x$ along the length of the link from the rotation axis.

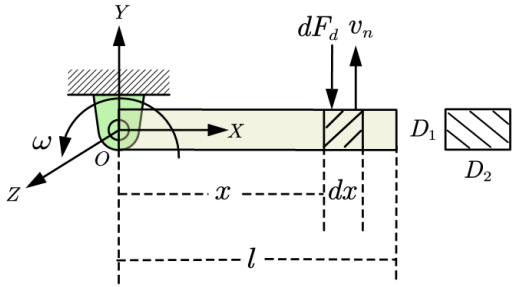

Figure 2 Hydrodynamic force calculation of a link

Assuming the normal velocity and water resistance are $v_n$ and $dF_d$, respectively, then the torque of the water resistance on the link with respect to the rotation axis $Z$ can be calculated by integrating the Morison formula along the length of the link, and the expression is:

$$\tau_d = \frac{1}{2}\rho C_d D_1 \int_0^l x v_n |v_n| dx \qquad (3)$$

Similarly, the torque with respect to the rotation axis $Z$ of the additional mass force on the link can be calculated through integration along the length of the link based on the Morison's formula:

$$\tau_m = \rho C_m A \int_0^l x \dot{v}_n dx \qquad (4)$$

where $A$ is the projected area of the link, $A = D_1 D_2$.

*A. Modeling of the hydrodynamic force of the UWML*

Assuming that the UWML works in a static water, the effect of current is ignored. Since the UWML has no wing-like structure, the effect of lift force $F_l$ can also be ignored. Therefore, the UWML is mainly disturbed by the water resistance $F_d$, the additional mass force $F_m$, and the buoyancy $F_f$. For convenience, the above three forces will be referred as hydrodynamic force of the UWML as a whole in the subsequent research of this paper. To investigate the influence of the hydrodynamic force on the dynamic characteristics of the UWML, it must be converted it into torque acting on the joints.

As shown in Fig. 3, a coordinate system is established for the analysis of the hydrodynamic force of the UWML, in which {1}~{3} represent the coordinate systems of joint 1~3, $q_1$~$q_3$ the joint angles, and $L_1$~$L_3$ the link lengths (since the shafts of joint 1 and joint 2 intersect, the link length $L_1 = 0$), respectively. For convenience, all the links are simplified as regular rectangular structures, and the centroid and the center of buoyancy of each link are assumed to be coincident but have opposite direction.

(1) The hydrodynamic torque modeling of joint 1

The hydrodynamic torque of joint 1 is generated from the interaction of link 2, link 3 and water during the motion. Based on the geometric relationship of the links shown in Fig. 3(b), a slice $dx$ is taken at a distance $x$ from each link's rotation axis in the direction of its length, and then the normal velocity of the slice on each link relative to the rotation axis $z_1$ are:

$$v_{11} = \omega_1 \times r_{11} \qquad (5)$$

$$v_{21} = \omega_1 \times r_{21} = \dot{q}_1 x \cos(q_2) \qquad (6)$$

$$v_{31} = \omega_1 \times r_{31} = \dot{q}_1 [L_2 \cos(q_2) + x \sin(q_2 + q_3)] \qquad (7)$$

where $\omega_1$ is the angular velocity of joint 1, and $r_{ij}$ is the distance between the cross-section on link $i$ and the origin of the coordinate system of joint $j$. Since joints 1 and 2 are designed integrally with intersecting axes, the distance $r_{11}$ can be regarded as zero if the influence of joint shape is ignored. The value of $r_{31}$ can be calculated based on the geometric relationship of the links, $r_{31} = \sqrt{L_2^2 + x^2 + 2L_2 x\sin(q_3)}$.

The water resistance torque $\tau_{d1}$ and additional mass force torque $\tau_{m1}$ of joint 1 can be calculated from the above velocity, respectively.

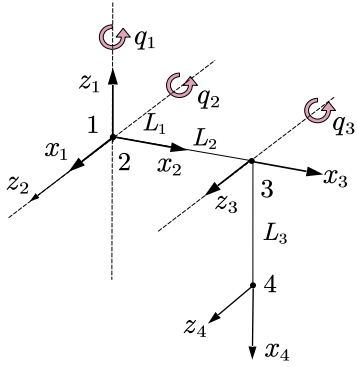
(a) Coordinate systems of the hydrodynamic force analysis

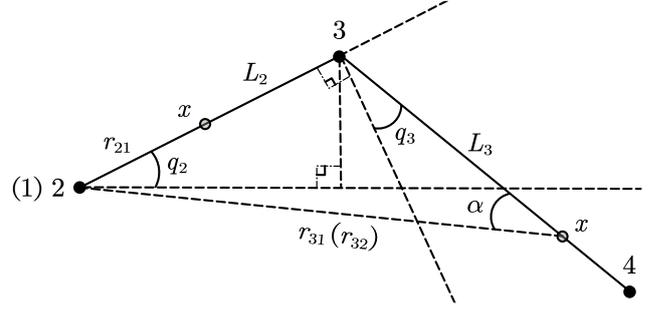
(b) Link geometry

Figure 3 Coordinate systems of the hydrodynamic force analysis of the UWML

$$\tau_{d1} = \frac{1}{2}\rho C_d D_{11}(\int_0^{D_1} x v_{11}|v_{11}|dx) + \frac{1}{2}\rho C_d D_{21}[\int_0^{L_2} x\cos(q_2)\, v_{21}|v_{21}|\, dx] \\ + \frac{1}{2}\rho C_d D_{31}\{\int_0^{L_3} [L_2\cos(q_2) + x\cos(q_2+q_3)]\, v_{31}|v_{31}|dx\} \quad (8)$$

$$\tau_{m1} = \rho C_m A_1(\int_0^{D_1} x v_{11}dx) + \rho C_m A_2[\int_0^{L_2} x\cos(q_2)\, v_{21}dx] \\ + \rho C_m A_3\{\int_0^{L_3} [L_2\cos(q_2) + x\sin(q_2+q_3)]\, v_{31}dx\} \quad (9)$$

where $D_{21}$ and $D_{31}$ are the effective height of the cross section of the link 2 and 3, respectively. $A_1 \sim A_3$ are the effective cross-sectional area of each link, which can be obtained from the link's effective height and width, namely, $A_1 = 0$, $A_2 = D_{21}D_{22}$, $A_3 = D_{31}D_{32}$. $D_{22}$ and $D_{32}$ are the effective widths of the cross sections of link 2 and 3, respectively.

As joint 1 is fixed on the frame, its axis is parallel to the direction of buoyancy, so the buoyancy has no torque effect on joint 1, namely:

$$\tau_{f1} = 0 \quad (10)$$

Assuming the direction of buoyancy torque acting on joint 1 is taken as the positive direction, then the total hydrodynamic torque of joint 1 can be calculated from all the hydrodynamic torque components as follows:

$$\tau_{w1} = \tau_{f1} - \tau_{d1} - \tau_{m1} \quad (11)$$

(2) The hydrodynamic torque modeling of joint 2

The hydrodynamic torque of joint 2 is caused by the interaction motion of link 2, link 3 and water, as well as the interaction between water and link 3 revolving around the axis $z_3$ of joint 3. Similarly, when joint 2 rotates around the axis $z_2$, the normal velocity of the slice on each link relative to the axis $z_2$ can be expressed as:

$$v_{22} = \omega_2 \times r_{22} = \dot{q}_2 x \quad (12)$$

$$v_{32} = \omega_2 \times r_{32} \\ = \dot{q}_2\sqrt{L_2^2 + x^2 + 2L_2 x\sin(q_3)} \quad (13)$$

where $r_{32} = r_{31}$ as the origins of coordinate system of joint 1 and 2 are coincide.

When link 3 rotates around the axis $z_3$ of joint 3, the normal velocity of the slice is:

$$v_{33} = \omega_3 \times r_{33} = \dot{q}_3 x \quad (14)$$

The water resistance torque $\tau_{d2}$ and additional mass force torque $\tau_{m2}$ of joint 2 can be calculated from the above velocity:

$$\tau_{d2} = \frac{1}{2}\rho C_d D_{22}(\int_0^{L_2} x v_{22}|v_{22}|dx) + \frac{1}{2}\rho C_d D_{32}[\int_0^{L_3} \sqrt{L_2^2 + x^2 + 2L_2\, x\sin(q_3)} \times v_{32}|v_{32}|dx] \\ + \frac{1}{2}\rho C_d D_{32}[\int_0^{L_3} \sqrt{L_2^2 + x^2 + 2L_2 x\sin(q_3)}\cos(\alpha)v_{33}|v_{33}|dx] \quad (15)$$

$$\tau_{m2} = \rho C_m A_2 \left( \int_0^{L_2} x \dot{v}_{22} dx \right) + \rho C_m A_3 \left[ \int_0^{L_3} \sqrt{L_2^2 + x^2 + 2L_2 x \sin(q_3)} \dot{v}_{32} dx \right]$$
$$+ \rho C_m A_3 \left[ \int_0^{L_3} \sqrt{L_2^2 + x^2 + 2L_2 x \sin(q_3)} \cos(\alpha) \dot{v}_{32} dx \right] \quad (16)$$

where $\alpha$ is the angle between $r_{32}$ and link 3, $\cos(\alpha) = (x^2 + r_{32}^2 - L_2^2)/(2xr_{32})$.

The buoyancy torque on joint 2 is generated from the buoyancy of link 2 and 3 acting on joint 2. Since the center of buoyancy of the link coincides with the center of gravity but in the opposite direction, the buoyancy torque $\tau_{f2}$ can be transformed into an expression about gravity:

$$\tau_{f2} = \{m_2 g L_{2c} \cos(q_2) + m_3 g [L_2 \cos(q_2) + L_{3c} \sin(q_2 + q_3)]\} \frac{\rho}{\rho_m} \quad (17)$$

where, $\rho$ and $\rho_m$ are the density of water and the equivalent density of the link, $L_{2c}$ and $L_{3c}$ the distance of the centroid of the link 2 and the link 3 relative to the axis of joint 2 and joint 3, respectively.

According to the components of the hydrodynamic torque calculated above, the total hydrodynamic torque on joint 2 can be obtained as follows:

$$\tau_{w2} = \tau_{f2} - \tau_{d2} - \tau_{m2} \quad (18)$$

(3) The hydrodynamic torque modeling of joint 3

The hydrodynamic torque on joint 3 is generated from the rotation of link 3 around joint 3, as well as the interaction between link 2 and link3. According to the normal velocities of the slices on link 3 with respect to the $z_2$ and $z_3$, namely $v_{32}$ and $v_{33}$, the water resistance torque $\tau_{d3}$ of joint 3 can be calculated as:

$$\tau_{d3} = \frac{1}{2} \rho C_d D_{32} \left( \int_0^{L_3} x v_{33} |v_{33}| dx \right)$$
$$+ \frac{1}{2} \rho C_d D_{32} \left[ \int_0^{L_3} x \cos(\alpha) v_{32} |v_{32}| dx \right] \quad (19)$$

Similarly, the additional mass torque $\tau_{m3}$ of joint 3 is:

$$\tau_{m3} = \rho C_m A_3 \left( \int_0^{L_3} x \dot{v}_{33} dx \right)$$
$$+ \rho C_m A_3 \left[ \int_0^{L_3} x \cos(\alpha) \dot{v}_{32} dx \right] \quad (20)$$

The buoyancy torque on joint 3 is produced by the buoyancy of link 3 acting on joint 3 during motion. Similarly, the buoyancy torque of joint 3 $\tau_{f3}$ can be transformed into an expression about gravity:

$$\tau_{f3} = m_3 g L_{3c} \sin(q_2 + q_3) \frac{\rho}{\rho_m} \quad (21)$$

According to the components of hydrodynamic torque calculated about joint 3, the total hydrodynamic torque on joint 3 can be obtained as follows:

$$\tau_{w3} = \tau_{f3} - \tau_{d3} - \tau_{m3} \quad (22)$$

*B. Fitting of the hydrodynamic parameters of the UWML*

Through the hydrodynamic modeling process described above, it can be seen that the drag force coefficient $C_d$ and the added mass force coefficient $C_m$ are two important parameters for calculation of the hydrodynamic torque of the UWML. However, the accurate values of these parameters cannot be obtained theoretically. Therefore, experimental data are used to fit the values of these parameters.

For convenience of testing, assumptions are made for the UWML: (1) When the UWML moving on land and in water with the same trajectory, it is assumed that the difference between the trajectories in the two environments under the same control algorithm and controller parameters will have a negligible effect on the joint output torque; (2) It is also assumed that the effect of the difference between the theoretical calculation results of the joint gravity torque and buoyancy torque and their actual values can be ignored; (3) The hydrodynamic parameters $C_d$ and $C_m$ are assumed to be constants.

Under the above assumptions, the testing process of the hydrodynamic parameters can be summarized into three steps: (1) The joint torques of the UWML tracking the same trajectory in both on land and underwater environments are tested separately, and the difference between the two is considered to be the hydrodynamic torque test results of each joint; (2) Based on the joint states (position, velocity, and acceleration) of the UWML and the parameters of the links, the hydrodynamic torque of each joint are calculated using the modeling results provided in Section 3.1; (3) The drag coefficient $C_d$ and the added mass force coefficient $C_m$ are determined through fitting of the hydrodynamic force test results with the theoretically calculated hydrodynamic model.

As the parameters of the links and the density of water are constant, and the joint states can be obtained through measurement or calculation (ignoring the influence of calculation errors), the drag force torque $\tau_d$ and additional mass force torque $\tau_m$ of each joint can be expressed as a linear function of $C_d$ and $C_m$, respectively:

$$\begin{cases} \tau_d = \alpha C_d \\ \tau_m = \beta C_m \end{cases} \quad (23)$$

where the coefficients $\alpha$ and $\beta$ are calculated based on the modeling result of the joint hydrodynamic torque, which are related to the states of joints and the parameters of links.

By substituting equation (23) into (11), (18) and (22), the joint hydrodynamics torque with respect to $C_d$ and $C_m$ can be obtained.

$$\tau_w = \tau_f - \alpha C_d - \beta C_m \quad (24)$$

where $\tau_f$, $\alpha$ and $\beta$ can be calculated based on the link's parameters and the states of the joints.

In theory, the values of $C_d$ and $C_m$ can be calculated with the hydrodynamic values of any two different states. However, due to the inaccuracy of the testing results and the discrepancy between theoretical and actual results, the hydrodynamic parameters computed from equation (24) may fluctuate greatly and become distorted. Therefore, this paper adopts the nonlinear fitting method to determine the values of hydrodynamic parameters.

The experimental platform used to test the hydrodynamic torque of the UWML is shown in Fig. 4, and its main components are listed in TABLE I. The same control method and parameters were used for the testing in both environments, and the foot trajectory was adopted as the one planned in reference [24], as shown in Fig. 5. The joint torques and the calculated joint hydrodynamic torques from the test are shown in Fig. 6, respectively. The parameters used for the calculation are shown in TABLE II.

According to Fig. 6(a), it can be seen that the torque difference of joint 2 is relatively large compared with joint 1 and joint 3. This is mainly due to the strong influence of gravity and buoyancy on the joint 2. By utilizing the results of Fig. 6(b) and the calculated values of $\alpha$ and $\beta$, the hydrodynamic parameters can be fitted according to equation (24). To improve the accuracy of the calculation and minimize the impact of errors in gravity and buoyancy on the test results, this paper utilized the test results of joint 1 for fitting (as the direction of gravity and buoyancy of joint 1 is always parallel to the axis of joint 1, they do not produce influence on joint 1). The fitting results of the hydrodynamic parameters are shown in Fig. 7. As seen, the values of $C_d$ and $C_m$ are about 2.2 and 0.5, and the fitting curve can effectively reflect the changing trend of the actual hydrodynamic force values (COD value reaches 0.826), which can be applied in engineering.

According to the fitted $C_d$ and $C_m$, the hydrodynamic torques can be calculated, which are presented in Fig. 7(c) with the actual results. It can be seen that although the hydrodynamic torque curves of the joint 2 and 3 are not matched very well, the calculated values can generally reflect the trend of the actual values but with a little bias relative to the actual values, especially for the joint 2. The reasons for this phenomenon mainly lie in the following aspects: (1) The simplification of the UWML structure for the calculation of the hydrodynamic torque neglects the influence of some complex shapes; (2) The differences of the response characteristics of the UWML under the two environments will have some impact on the joint torques, which will be directly reflected in the measured hydrodynamic torques; (3) The inaccurate calculation of buoyancy and gravity for each link, as well as the non-coincidence of the centers of buoyancy and gravity, can also affect the test results of the hydrodynamic torques. (4) With different motion states, the interaction of the UWML with water is also different, which may cause changes in the water flow state relative to the UWML (such as from laminar flow to turbulent flow, or vortex etc.), thereby results in changes of $C_d$ and $C_m$, which will in turn affect the test results of the hydrodynamic torques. hydrodynamic torques.

### C. Study on the characteristics of the hydrodynamic force of the UWML

Hydrodynamic force, as a disturbance, will affect the dynamic performance of the UWML. Research on the hydrodynamic characteristics of the UWML can deeply analyze the degree of influence of each component of the hydrodynamic force on the dynamic performance of the UWML. Since the hydrodynamic force components are acting on the UWML as a whole, it is impossible to test the change laws of any single component through experiment. Therefore, this paper utilized the hydrodynamic model with the fitted parameters to study the change laws of the hydrodynamic force on each joint of the UWML. The results are shown in Fig. 8.

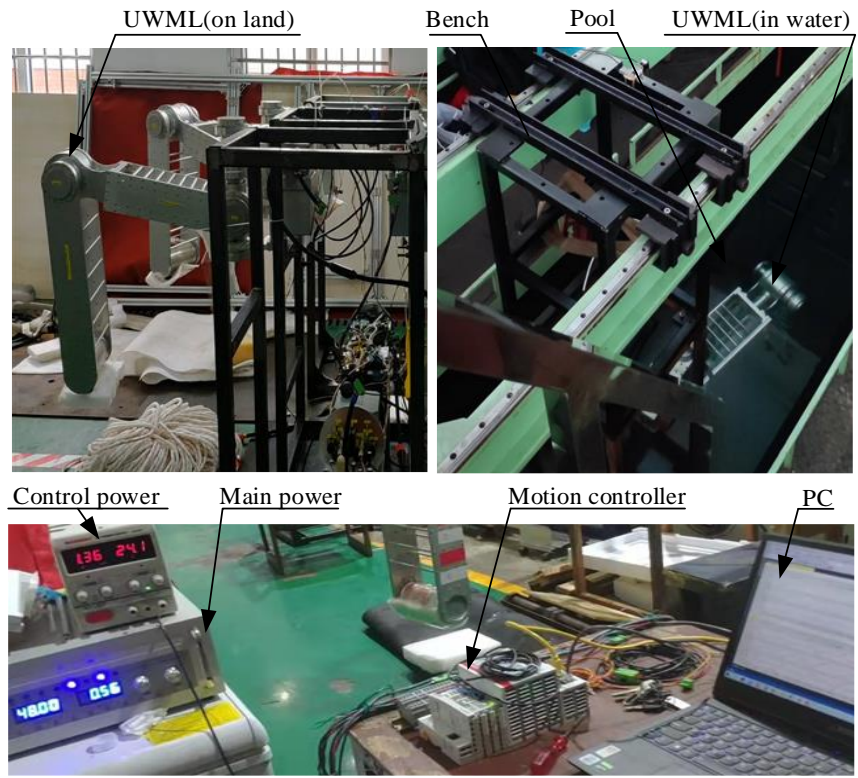

Figure 4 Hydrodynamic force testing platform for the UWML

TABLE I. MAIN COMPONENTS OF THE HYDRODYNAMIC FORCE TESTING PLATFORM OF THE UWML

| Name | Specification | Name | Sspecification |
|---|---|---|---|
| Motor for joint 1/2/3 | Kollmorgen TBMS-7646-A | Velocity sensor of joint 2 | Tamagawa TS2620N271E14 |
| reducer for joint 1/2/3 | Dongguan Benyun BHS32-160 | Position sensor of joint 3 | Tamagawa TS2640N321E64 |
| Motor driver for joint 1/2/3 | Elmo G-SOLTWI15/100ER1 | Velocity sensor of joint 3 | Tamagawa TS2620N271E14 |
| Position sensor of joint 1 | Tamagawa TS2660N31E148 | Motion controller | Beckhoff CX5140 PLC |
| Velocity sensor of joint 1 | Tamagawa TS2620N271E14 | Main power | DC48V |
| Position sensor of joint 2 | Tamagawa TS2620N271E14 | Control power | DC24V |

TABLE II. PARAMETERS OF THE UWML

| Link length (m) | COM of link (m) | Section size of link (m) | Link mass (kg) | Density of link/water (kg/m³) |
|---|---|---|---|---|
| $L_1 = 0$ | $L_{c1} = 0$ | $D_{11} = 0, D_{12} = 0$ | $m_1 = 10.758$ | $\rho_m = 2700$ $\rho = 1000$ |
| $L_2 = 0.660$ | $L_{c2} = 0.506$ | $D_{21} = 0.131, D_{22} = 0.238$ | $m_2 = 19.261$ | |
| $L_3 = 0.714$ | $L_{c3} = 0.560$ | $D_{31} = 0.131, D_{32} = 0.279$ | $m_3 = 10.375$ | |

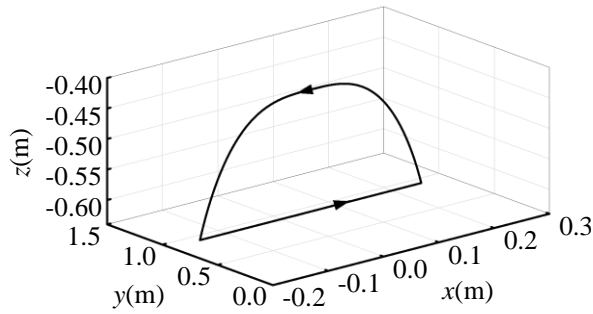

Figure 5 Foot trajectory of the UWML

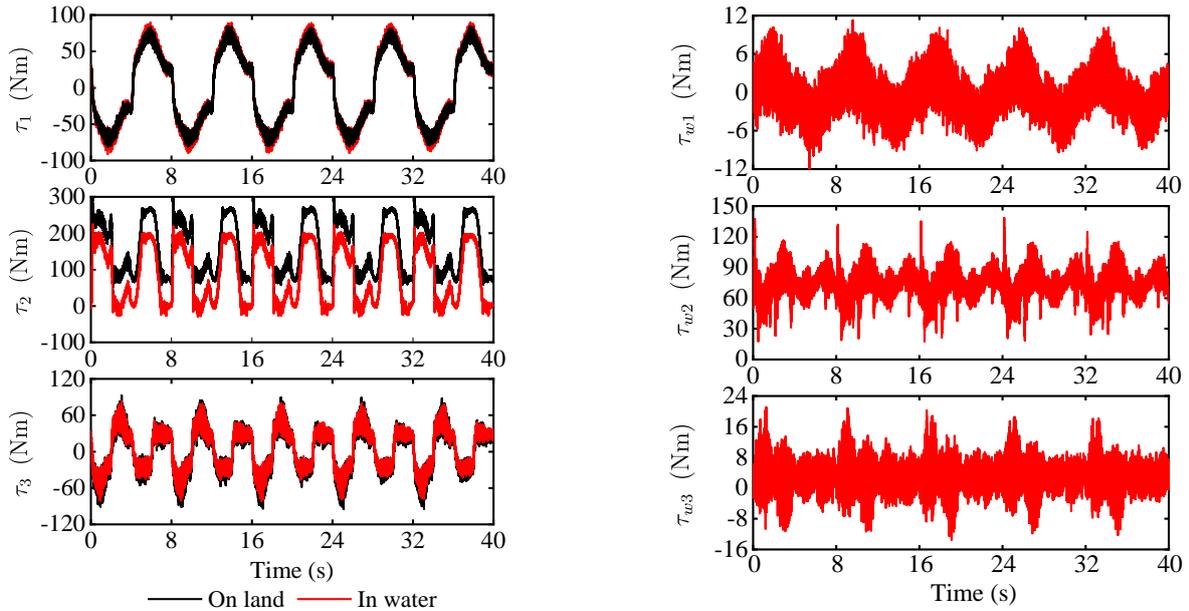

(a) Joint torques on land /in water

(b) Hydrodynamic torques of the joints

Figure 6 Hydrodynamic force testing results of the UWML

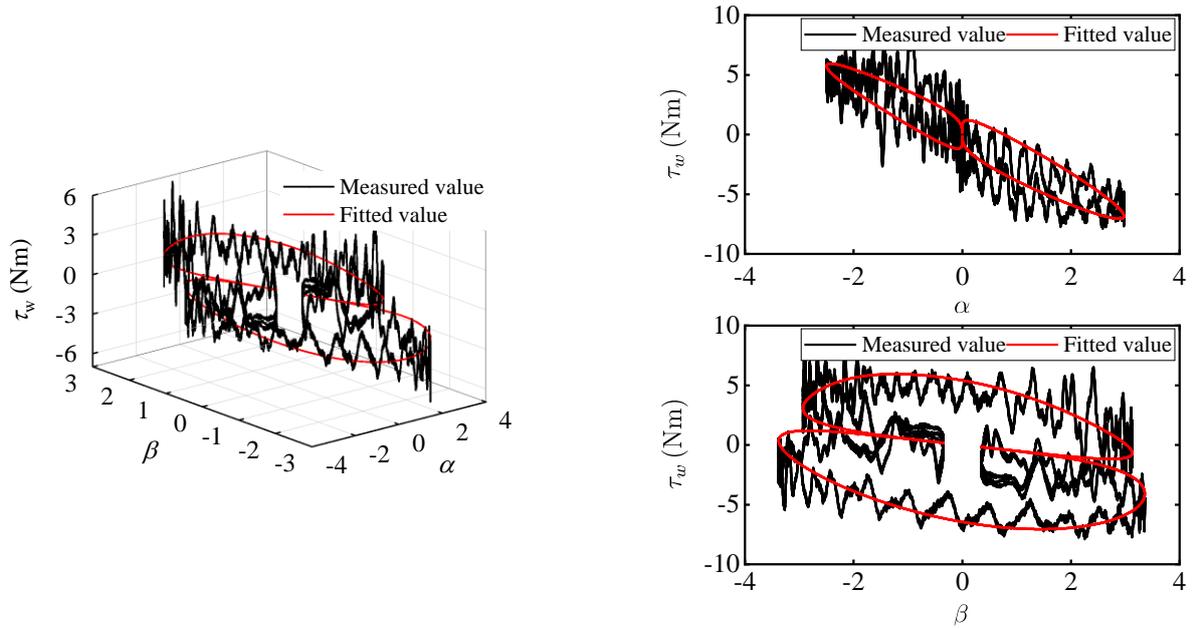

(a) Hydrodynamic fitting curves of joint 1

| Fitted results | |
|---|---|
| Function | τw-τf = -α×Cd-β×Cm |
| Plot | τw-α, τw-β |
| Cd | 2.19747 ±0.00551 |
| Cm | 0.50025 ±0.00344 |
| Reduced Chi-Sqr | 2.425 |
| COD | 0.82605 |

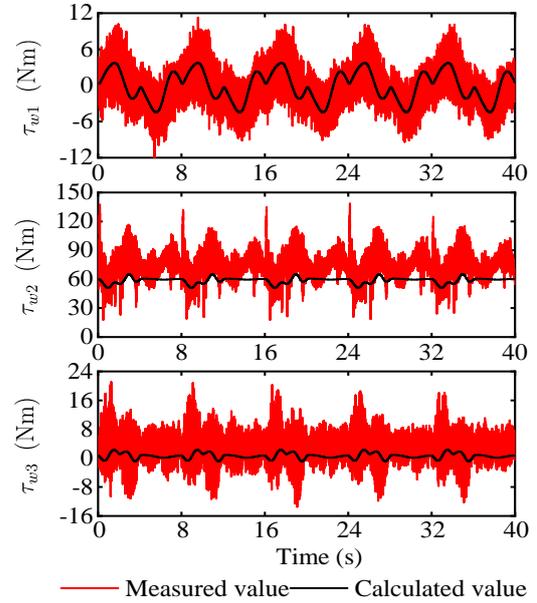

(b) Fitting results of the hydrodynamic parameters  (c) Comparison results of the calculated and measured hydrodynamic torques

Figure 7 Fitting results of the hydrodynamic parameters of the UWML

It can be seen from Fig. 8 that the hydrodynamic torques and their components vary with joint states with the same motion period of the UWML. In addition, by comparing the magnitudes of the drag force torque $\tau_d$, the additional mass force torque $\tau_m$, and the buoyancy torque $\tau_f$ of each joint in Fig. 8(b) ~ (d), it can be concluded that $\tau_f$ is the largest, followed by $\tau_d$, and $\tau_m$ is the smallest, and the values of $\tau_d$ and $\tau_m$ are much smaller than $\tau_f$, especially for $\tau_m$.

## IV. RESEARCH ON THE CHARACTERISTICS OF VISCOUS RESISTANCE AND DYNAMIC SEAL RESISTANCE OF THE WATERTIGHT JOINT

Due to oil compensation and dynamic seal, the joints of the UWML are subjected to viscous resistance and dynamic seal resistance when working in deep-sea environments. Although researchers have summarized some empirical formulas to describe the variation laws of both resistance, there are many parameters involved, and lack of generality [8-11]. Therefore, experimental test is commonly used for the research in practice. This section analysis the change laws of the viscous resistance and dynamic seal resistance of a watertight joint with respect to rotational speed and environmental pressure under different conditions.

### A. Watertight joint testing system

The watertight joint testing system was constructed using the hip roll joint of the UWML, as shown in Fig. 9, which is mainly composed of the joint and a pressure test system. Three sets of experiments were conducted under different working conditions, from which the change laws of the viscous resistance and output shaft dynamic seal resistance can be obtained indirectly.

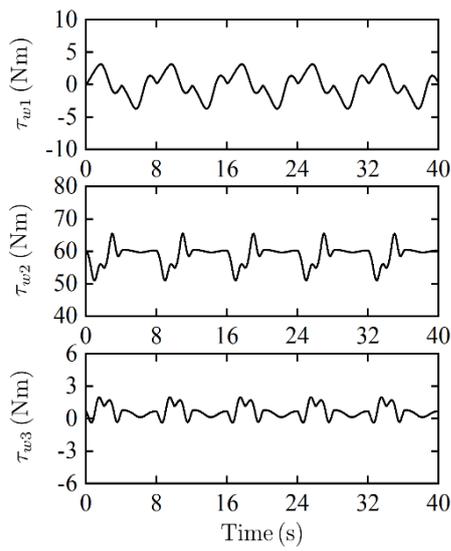
(a) Hydrodynamic torque of joints

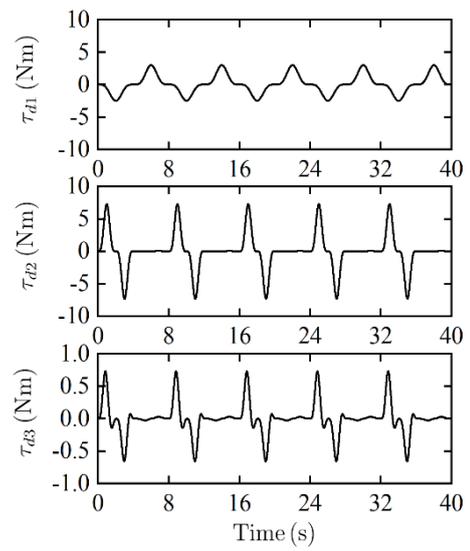
(b) Drag force torque of joints

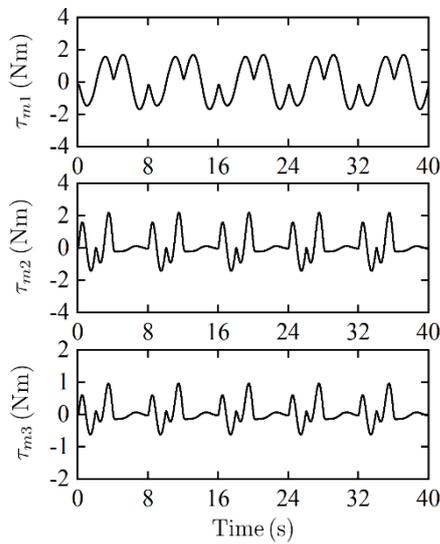
(c) Added mass force torque of joints

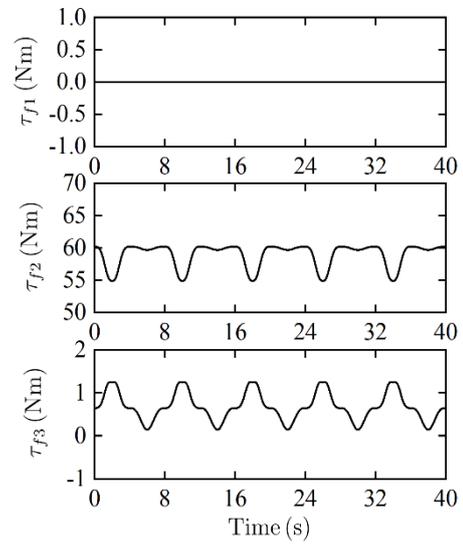
(d) Buoyance torque of joints

Figure 8 The curve of the hydrodynamic torque and its component

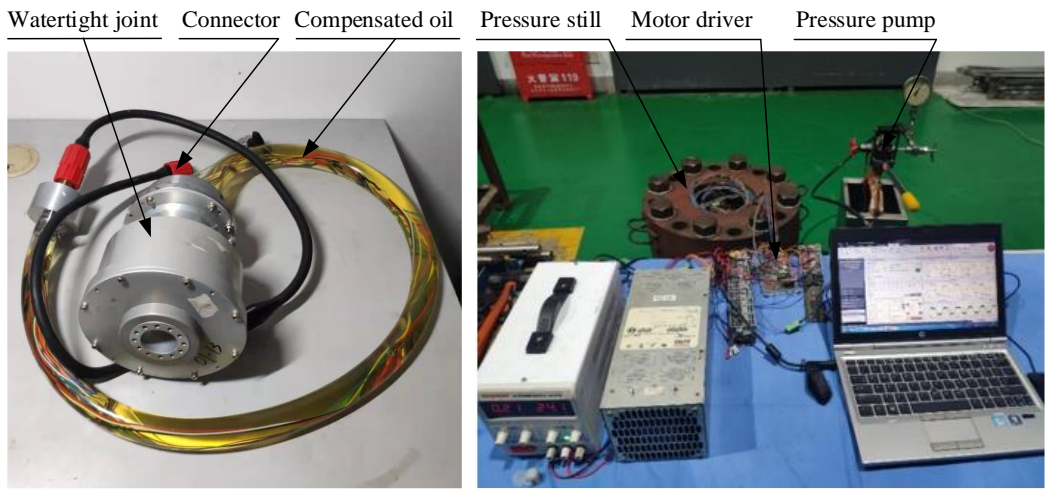

(a) Watertight joint.    (b) Pressure testing system.

Figure 9 Watertight joint testing system

## B. Characteristics study of the viscous resistance and dynamic seal resistance

Three sets of experiments were conducted with the watertight joint. The first set involved performance testing of the joint at different rotation speed but with no oil filled in the housing cavity and dynamic sealing components installed on the output shaft, while the second set with the joint in different rotation speeds and working pressure in oil-filled the housing cavity and dynamic sealing components installed on the output shaft, while the second set with the joint in different rotation speeds and working pressure in oil-filled but with no dynamic seal components installed, and the third set with the joint in different rotation speeds and working pressure in oil-filled and with dynamic seal components installed. The test data for the experiments are presented in Fig. 10 to Fig. 12.

The change laws of the working current corresponding to the viscous resistance under different rotation speed and pressure can be obtained by comparing the results of Fig. 10 and Fig. 11, as shown in Fig. 13. Similarly, the change laws of the working current corresponding to the dynamic seal resistance under different rotation speeds and pressure can be obtained from the results of Fig. 11 and Fig. 12, as shown in Fig. 14. As there is a constant relationship between the working current and the viscous resistance and the dynamic seal resistance, the changing laws of the corresponding working current also reflect the change laws of the resistance. Therefore, the following aspects directly describe the viscous resistance and the dynamic seal resistance with the comparing results.

The results of Fig. 13 and Fig. 14 indicate that both viscous resistance and dynamic seal resistance increase with the increasing of speed and pressure. However, their sensitivities to changes in speed and pressure are different, with viscous resistance being more sensitive to changes in speed, while dynamic seal resistance being more sensitive to changes in pressure. Additionally, it also can be seen that, under high-pressure conditions, the dynamic seal resistance is also very high during the startup of the joint motion.

Further analysis of the results in Fig. 12 reveals that the total current loss, caused by the viscous resistance and dynamic seal resistance at a high speed of 10r/min and high pressure of 10MPa, is approximately 2.85A, almost 27.4% of the motor rated current (10.4A). This means that the efficiency of the joint operating in a kilometer-level deep-sea environment is only 72.6% compared with that on land, about 27.4% of energy lost due to the two resistance.

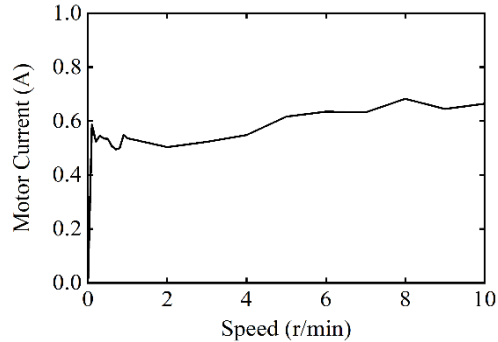

Figure 10 Motor current in different rotation speeds with no oil filled in housing cavity and dynamic sealing components installed

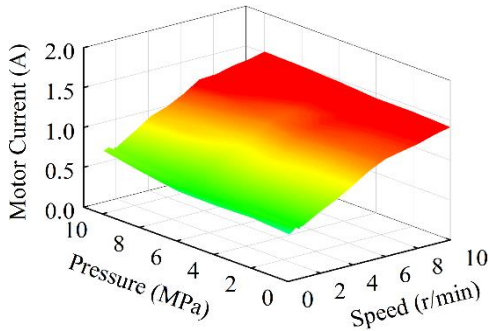

Figure 11 Motor current in different rotation speed and pressure with oil filled but without dynamic seal components installed

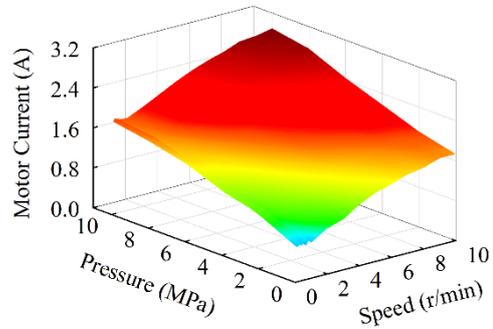

Figure 12 Motor current in different rotation speed and pressure with oil filled and dynamic seal components installed

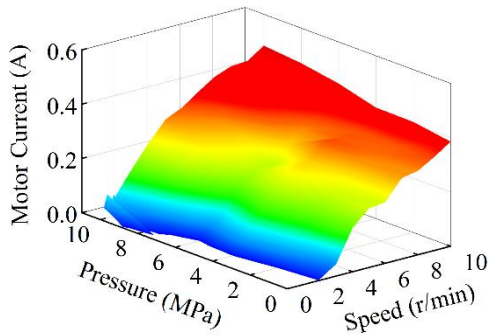

Figure 13 Oil viscous resistance in different rotation speed and pressure

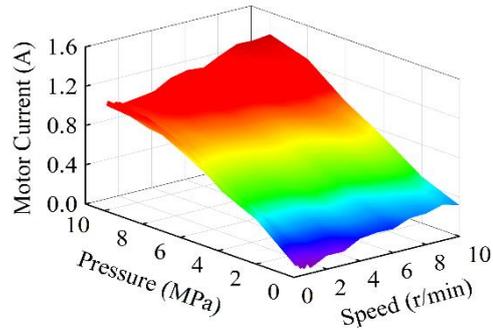

Figure 14 Dynamic seal resistance in different rotation speed and pressure

The above test results for watertight joint demonstrate that, when the UWML operates in a deep-sea environment, joint viscous resistance and dynamic seal resistance cause a significant power loss, which will further affect the system's dynamic performance, and the effects of both will be aggravated with the increase of depth. Subsequent research on the design of high performance controllers for the UWML must consider the influence of the viscous resistance and dynamic seal resistance to reduce their impact on control performance.

## V. CONCLUSION

This paper investigated the influence of the underwater environment on the dynamic performance of the UWML of the DCSR. The hydrodynamic model was established with the fitted parameters, and the change rules of the viscous resistance and dynamic seal resistance of the joint were studied:

(1) For the modeling of the hydrodynamic force of the UWML, experiments were conducted indirectly, and the hydrodynamic parameters were obtained by nonlinear fitting method with the theoretical model and the test data, which are $C_d \approx 2.2$ and $C_m \approx 0.5$. Then the model of the hydrodynamic force was established with the fitted parameters. Furthermore, the proportion of each hydrodynamic force component was analyzed with the hydrodynamic model. The results show that the buoyancy torque is the largest, the additional mass torque the least, and the water resistance torque is somewhere in between.

(2) For the impact of viscous resistance and dynamic seal resistance on the joint of the UWML, the test experiments were conducted under three different conditions. The results show that the viscous resistance and dynamic seal resistance of the joint increase with the increasing of the rotation speed and working pressure, but their sensitivity to the change of rotational speed and working pressure is not the same, namely, the viscous resistance is more sensitive to the change of speed, but the dynamic seal resistance is more sensitive to the change of pressure. Furthermore, the efficiency of the joint was analyzed, and the results indicate that about 27.4% of the efficiency will lost in the kilometer-level deep-sea environment due to the effects of viscous resistance and dynamic seal resistance.

The research of the change laws of the hydrodynamic force, viscous resistance and dynamic seal resistance of the UWML is of great significance. The results will provide important guidance in optimizing the structure of the next generation of the UWML, improving the control performance of the UWML for the crawling and underwater manipulating. This research also lays a foundation for the designing of an electric underwater manipulator for the deep sea operation.

## ACKNOWLEDGMENT

This work was supported by the National Key Research and Development Program of China (Grant No. 2016YFC0301700), which is of great appreciation.